\documentclass[sigconf]{acmart}
\usepackage{multirow}
\usepackage{stfloats}
\usepackage{enumitem}
\usepackage{fancyvrb}

\usepackage{lipsum}

\AtBeginDocument{%
  \providecommand\BibTeX{{%
    \normalfont B\kern-0.5em{\scshape i\kern-0.25em b}\kern-0.8em\TeX}}}

\setcopyright{acmcopyright}
\copyrightyear{2018}
\acmYear{2018}
\acmDOI{XXXXXXX.XXXXXXX}

\acmConference[Conference acronym 'XX]{Make sure to enter the correct
  conference title from your rights confirmation emai}{June 03--05,
  2018}{Woodstock, NY}
\acmPrice{15.00}
\acmISBN{978-1-4503-XXXX-X/18/06}

\begin{document}

\title{ID Embedding as Subtle Features of Content and Structure for Multimodal Recommendation}


\author{Yuting Liu}
\authornote{Both authors contributed equally to this research.}
\email{yutingliu@stumail.neu.edu.cn}
\author{Enneng Yang}
\authornotemark[1]
\email{ennengyang@stumail.neu.edu.cn}
\affiliation{%
  \institution{Northeastern University}
  \country{China}
}

\author{Yizhou Dang}
\affiliation{%
  \institution{Northeastern University}
  \country{China}
}
\email{dangyz@stumail.neu.edu.cn}

\author{Guibing Guo}
\authornote{Corresponding author.}
\affiliation{%
  \institution{Northeastern University}
  \country{China}
}
\email{guogb@swc.neu.edu.cn}

\author{Qiang Liu}
\affiliation{%
  \institution{Chinese Academy of Sciences}
  \country{China}
}
\email{qiang.liu@nlpr.ia.ac.cn}

\author{Yuliang Liang}
\affiliation{%
  \institution{Northeastern University}
  \country{China}
}
\email{liangyuliang@stumail.neu.edu.cn}

\author{Linying	Jiang}
\affiliation{%
  \institution{Northeastern University}
  \country{China}
}
\email{jiangly@swc.neu.edu.cn}

\author{Xingwei Wang}
\affiliation{%
 \institution{Northeastern University}
 \country{China}
}
\email{wangxw@swc.neu.edu.cn}

\renewcommand{\shortauthors}{Yuting and Enneng, et al.}

\begin{abstract}
  Multimodal recommendation aims to model user and item representations comprehensively with the involvement of multimedia content for effective recommendations. Existing research has shown that it is beneficial for recommendation performance to combine (user- and item-) ID embeddings with multimodal salient features, indicating the value of IDs. However, there is a lack of a thorough analysis of the ID embeddings in terms of feature semantics in the literature. In this paper, we revisit the value of ID embeddings for multimodal recommendation and conduct a thorough study regarding its semantics, which we recognize as subtle features of \emph{content} and \emph{structure}. Based on our findings, we propose a novel recommendation model by incorporating ID embeddings to enhance the salient features of both content and structure. Specifically, we put forward a hierarchical attention mechanism to incorporate ID embeddings in modality fusing, coupled with contrastive learning, to enhance content representations. Meanwhile, we propose a lightweight graph convolution network for each modality to amalgamate neighborhood and ID embeddings for improving structural representations. Finally, the content and structure representations are combined to form the ultimate item embedding for recommendation. Extensive experiments on three real-world datasets (Baby, Sports, and Clothing)  demonstrate the superiority of our method over state-of-the-art multimodal recommendation methods and the effectiveness of fine-grained ID embeddings. Our code is available at \url{https://anonymous.4open.science/r/IDSF-code/}.
\end{abstract}





\maketitle

\section{Introduction}
Research on multimodal recommendation has flourished, primarily thanks to its capability to improve item representations through integrating various sources of multimedia information~\cite{Zhou2020S3RecSL,He2018AdversarialPR,Cao2019UnifyingKG,incorporatingLiu2022},  with text and image being the two most widely adopted modalities.
A prevalent viewpoint in existing research is that the key to enhance item representations is to extract salient semantic features from multiple modalities for effective item representations. 
In contrast, ID embedding has not attracted sufficient attention and has not been thoroughly explored in multimodal recommendation despite its well-demonstrated value in both traditional~\cite{mf,TrustSVDCF} and multimodal~\cite{vbpr,mmgcn,slmrec} recommendation. 
The specific information captured by ID embeddings remains unclear, resulting in suboptimal strategies for its utilization and leaving ample room for further improvements in multimodal recommendation. 

In this paper, we revisit the efficiency of ID embeddings and conduct case studies to explore the intricate details within them, which we refer to as subtle features.
We argue that ID embeddings should be regarded as a mixture of subtle features of both content and structure, as shown in Section~\ref{sec:analysis}. 
Based on the analysis, we classify the use of ID embeddings in the recommendation into two lines of research. Firstly, ID embeddings are regarded as vectors of \emph{content features}, containing some semantic attributes of an item. They are learned from the historical interactions between users and items via approaches such as matrix factorization~\cite{mf,dtmf}, and can be concatenated with multimodal features to enrich item representations in multimodal recommendation~\cite{vbpr,acf,deepstyle}. 
Secondly, ID embeddings are explained as vectors of \emph{structural features}, capturing the features of all the neighboring items. ID embeddings are trained from the user-item interaction bipartite graph, where each item is linked with a set of users (i.e., neighbors), through Graph Convolutional Networks (GCN)~\cite{Gori2007ItemRankAR, Bruna2014SpectralNA, Defferrard2016ConvolutionalNN} approaches. They can enhance item representations by being aggregated with each modal feature obtained from modal-specific interaction graphs in the multimodal recommendation.~\cite{mmgcn,grcn,slmrec, session_based_GNN, ultragcn, Liu2021PretrainingGT, Wu2021SelfsupervisedGL}.

\begin{figure}[t]
    \centering
    \includegraphics[width=\linewidth]{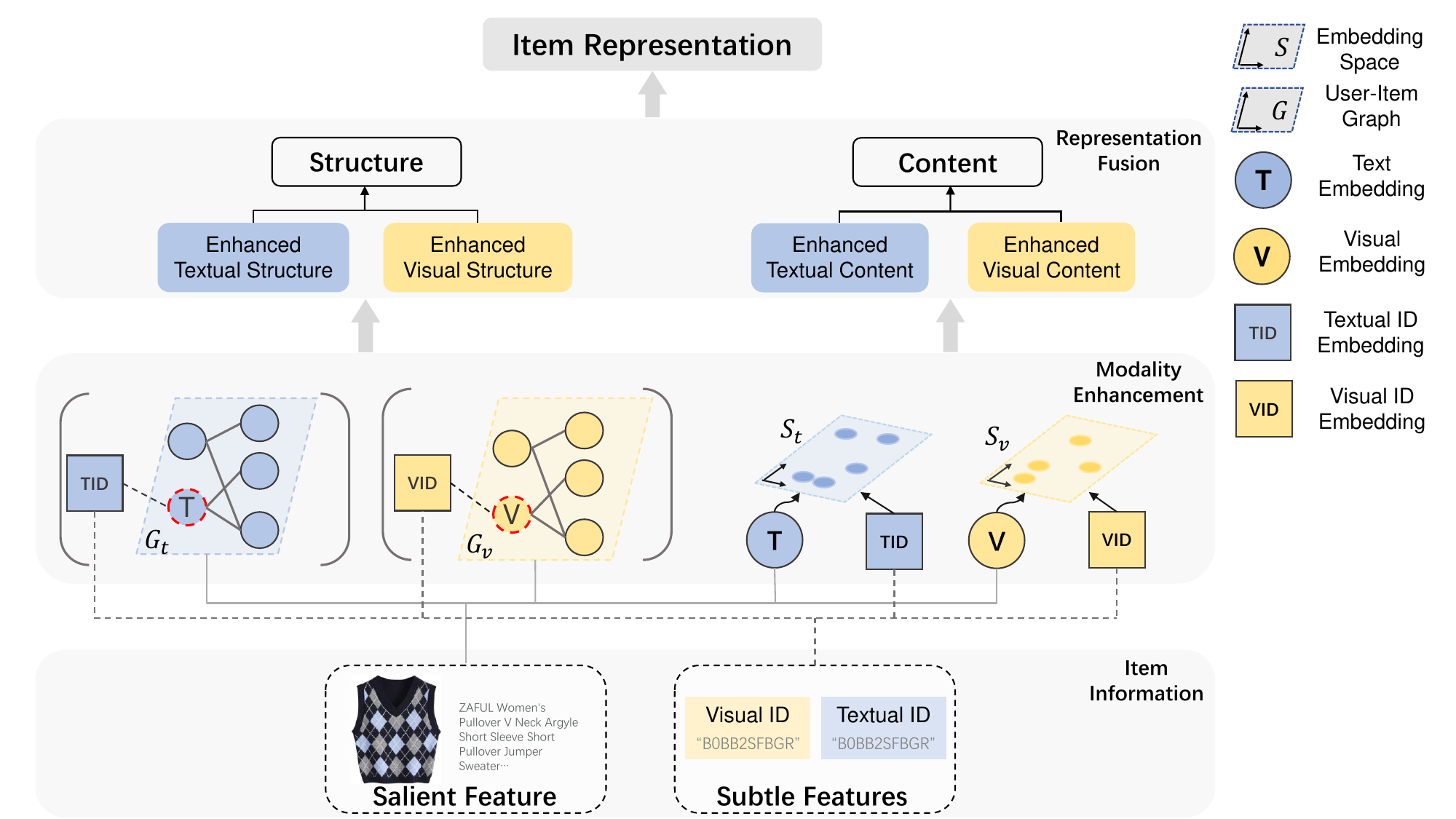}
    \vspace{-10pt}
    \caption{
    An item consists of textual and visual modalities and ID information. Features in terms of content and structure can be enhanced in each modality separately. The item representation can be obtained by fusing content and structural features. Note that traditional ID embeddings are treated as content or structural features, while our approach takes ID embeddings as a mixture of subtle features of both content and structure.
    }
    \label{fig:example}
    \vspace{-10pt}
\end{figure}

However, these approaches mainly focus on the salient features within multimodal sources and treat ID as a whole, combining it with them in a rudimentary manner. Consequently, most research concentrates on extracting and integrating multimodal information without the detailed exploration of content and structural features within ID. In our viewpoint, ID embeddings reflect the subtle features of items, which usage should be carefully considered from the standpoint of content and structure as an independent information source to enhance item representations.

On this basis, we propose a novel multimodal recommender called \textbf{IDSF} that explains \underline{ID} embeddings as \underline{S}ubtle \underline{F}eatures of both content and structure. It provides additional signals to enhance the semantics of extracted salient features under each modality regarding content and structure, leading to an improved item representation. 
Specifically, IDSF consists of two main modules to learn item content and structural features from all modalities. 
Figure~\ref{fig:example} illustrates an intuitive example derived from the \textit{Clothing}~\footnote{\footnotesize \url{https://cseweb.ucsd.edu/~jmcauley/datasets/amazon/links.html}} dataset, where an item consists of content and structural features learned from three sources of information (i.e. textual, visual, and item ID). We hierarchically integrate all the information to obtain a better comprehension of the item's semantics. For content features, we first enhance the salient features from textual and visual content by fusing them with the subtle features from modal-specific ID embeddings. Then, the enhanced modalities are fused through an attention mechanism and contrastive learning. Similarly, we retrieve the structural features by merging salient neighborhood and subtle features from a modality-specific interaction graph. Notes that we introduce modal-specific ID embeddings to help avoid mutual interference among different modalities which is proven to be necessary in Section~\ref{sec:analysis}. 
Finally, the item representation is obtained by combining the content and structural features.

It is worth mentioning that our method enjoys an additional benefit in alleviating the modal missing problem in the multimodal recommendation. 
That is, some modalities may be unavailable in real applications, leading to a decline in the performance of the existing multimodal recommenders~\cite{DeepAdversarialForMMM,LearningFactorized,DeepLearningBasedImaging,Cui2020MVRNNAM,WangLRMM,ZhouMultimodalBehavioral}. Our work can alleviate this problem by making use of subtle features of content and structure derived from ID embeddings and thus maintaining relatively high performance.

\ \\ \noindent \textbf{Contributions.} To sum up, the main contributions of our work are presented as follows: 
\begin{itemize}
    \item We highlight the significance of ID embeddings and undertake a thorough analysis to explore the optimal approach to leveraging ID embeddings in multimodal recommendations. To the best of our knowledge, we are making the initial endeavor to thoroughly explore the detailed information and employ fine-grained utilization with ID embeddings.
    \item We propose a novel framework that considers both item ID and modality in terms of content and structure to acquire comprehensive item representations. For content, we design a hierarchical attention mechanism and employ contrastive learning to enhance salient modal features with subtle content features within ID embeddings. For structure, we develop a modal-specific lightweight graph convolution network to involve subtle structural features in salient features. Eventually, we combine the content and structural features to obtain final item representations.
    \item We perform extensive experiments on three real-world datasets (Baby, Sports, and Clothing) to demonstrate that our method outperforms the state-of-the-art recommendation methods. In addition, we find that our approach can effectively mitigate the modal missing problem. 
\end{itemize}

\begin{figure}[t]
    \centering
    \includegraphics[scale=0.19]{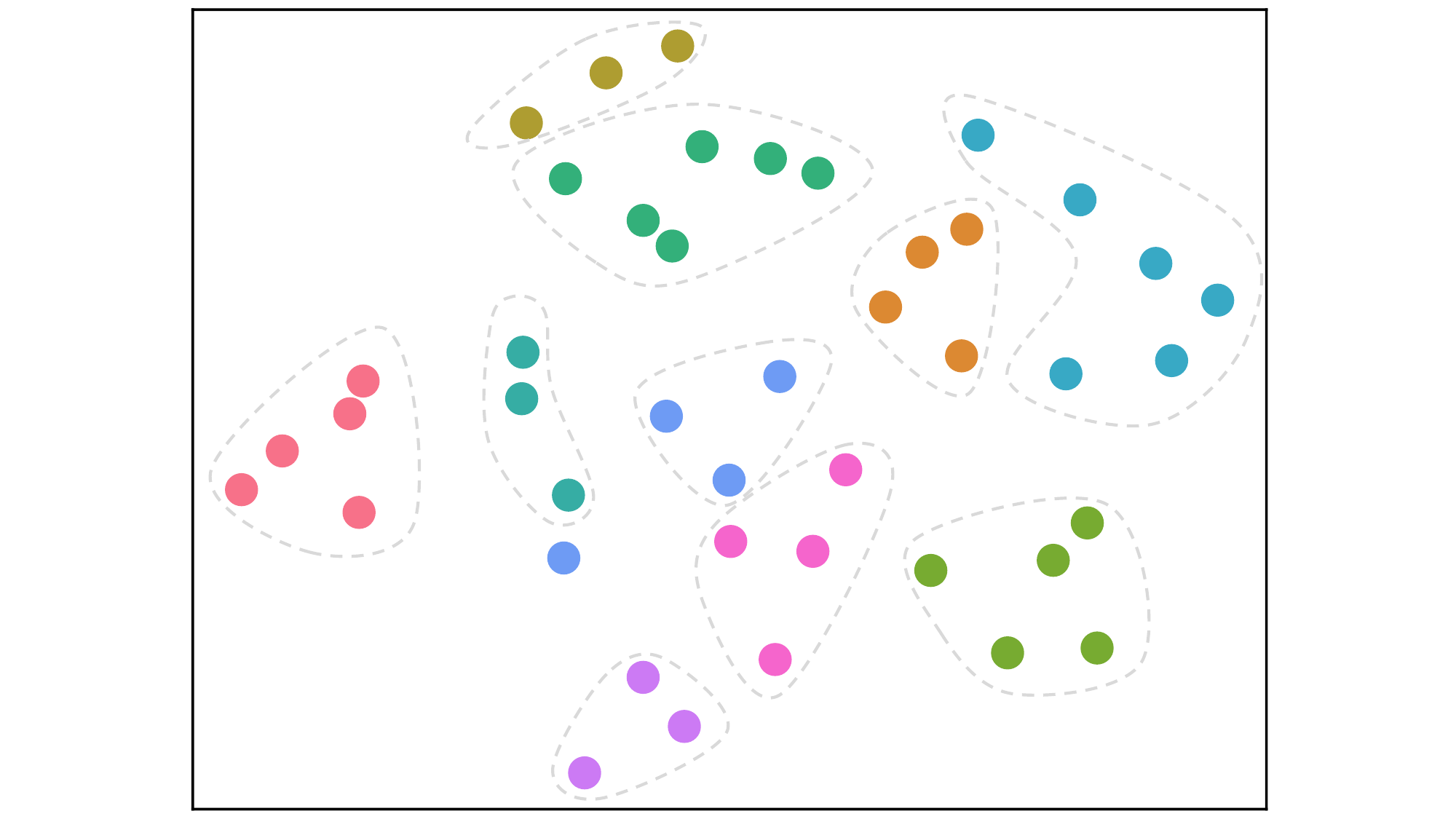}
    \vspace{-10pt}
    \caption{Distributions of pre-trained ID embeddings. The same color represents items that interacted with the same user.}
    \label{fig:id_scatter}
    \vspace{-10pt}
\end{figure}

\begin{figure}[t]
    \centering
    \includegraphics[scale=0.24]{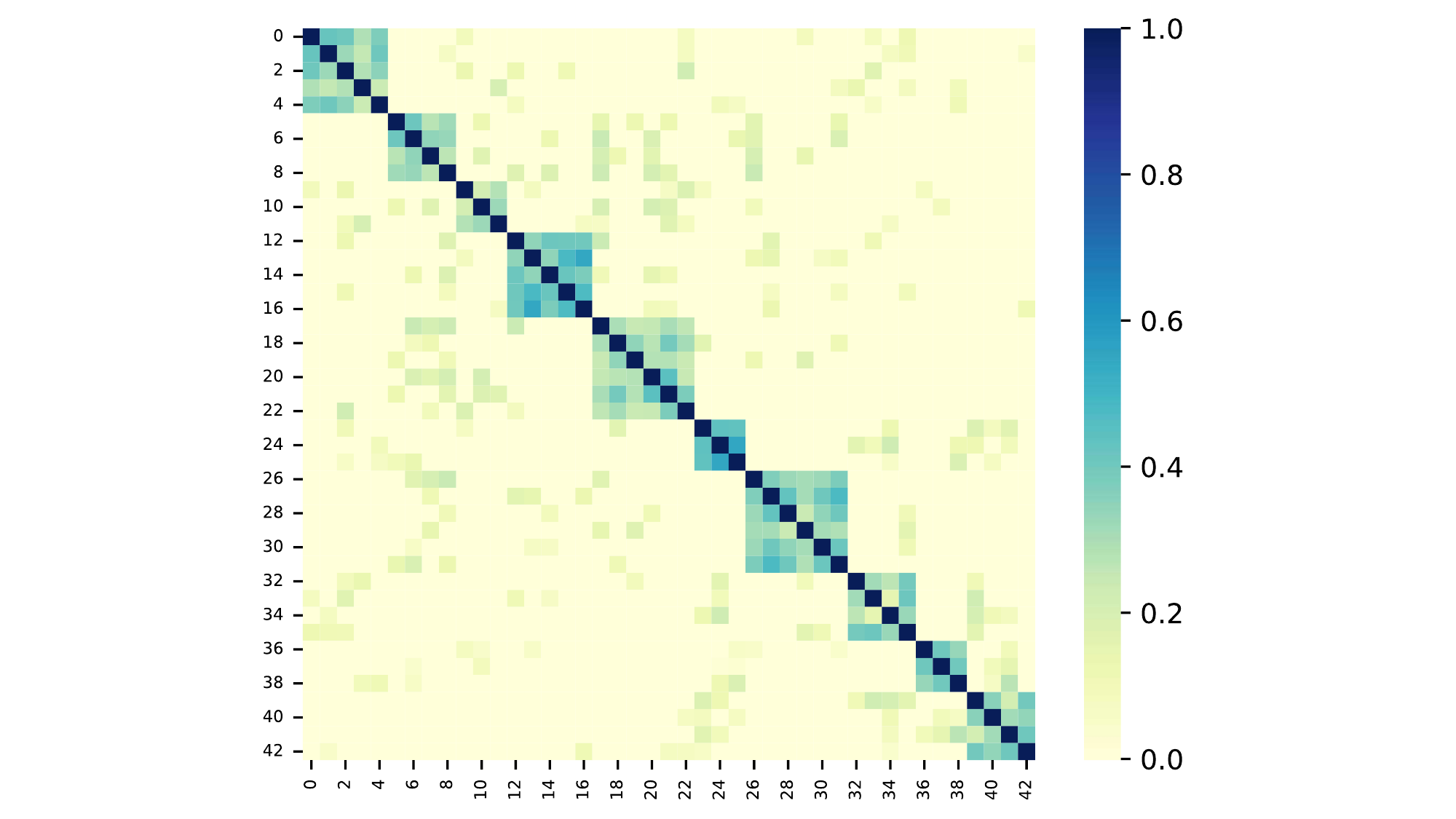}
    \vspace{-10pt}
    \caption{Semantic similarity of ID embeddings. Each cell in the heat map represents the normalized similarity of two ID embeddings, with horizontal and vertical coordinates denoting their mapped ID. Darker hues indicate a higher similarity.}
    \label{fig:id_heatmap}
    \vspace{-10pt}
\end{figure}

\section{ID Embedding Contains More}
\label{sec:analysis}

Yuan et al.~\cite{IDvsModal} have identified that while modal-based recommendation has achieved comparable performance to ID-based recommendation due to the advances in the multimodal domain, ID dominates recommendation with typical architecture. However, the underlying reasons for the remarkable effectiveness of the ID are still not fully understood. To delve deeper into this matter, we employed visualizations to analyze the ID embeddings and modal features, leading us to draw preliminary conclusions: the subtle features conveyed by ID embeddings, which consist of content and structural features, are anticipated to be modal-specific. Meanwhile, they can enhance the corresponding original modal features.


We first map the pre-trained item ID embeddings\footnote{We provide more details in Appendix~\ref{app:detail}.} to 2-dim normalized vectors by using UMAP~\cite{UMAP} and plot their distributions (shown in Figure~\ref{fig:id_scatter}). Note that points of the same color correspond to items that interacted with the same user. 
In Figure~\ref{fig:id_scatter}, the points of items interacted with different users form distinct clusters that are clearly visible and distinguished by the colors.  
This observation indicates that item ID embeddings of the same user have a closer distribution which is aligned with the edge in the user-item interaction graph.
This suggests that item ID embeddings capture \textit{structural features} within the user-item interaction graph.
Simultaneously, we generate a heat map to visualize the semantic similarity among these item ID embeddings (shown in Figure~\ref{fig:id_heatmap}). Each cell in the heat map represents the normalized similarity of two items, where the horizontal and vertical coordinates indicate their respective re-indexes. Note that we set adjacent coordinates for items that interact with the same user (e.g., indexes 0-4 correspond to the items that interacted with the same user). 
To improve clarity, we assign all values in the similarity matrix, except for the top 10 values in each row, to 0.
In Figure~\ref{fig:id_heatmap}, items that interacted with the same user exhibit a higher similarity and items interacted with different users also display certain semantic similarities due to the inherent content features of items themselves.
According to the observation, we posit that ID embeddings capture \textit{content features} of items.
Therefore, to enhance the utilization of ID embeddings in the multimodal recommendation, we suggest \emph{adopting a comprehensive approach that integrates content and structure concurrently}.

\begin{figure}
    \centering
    \includegraphics[scale=0.24]{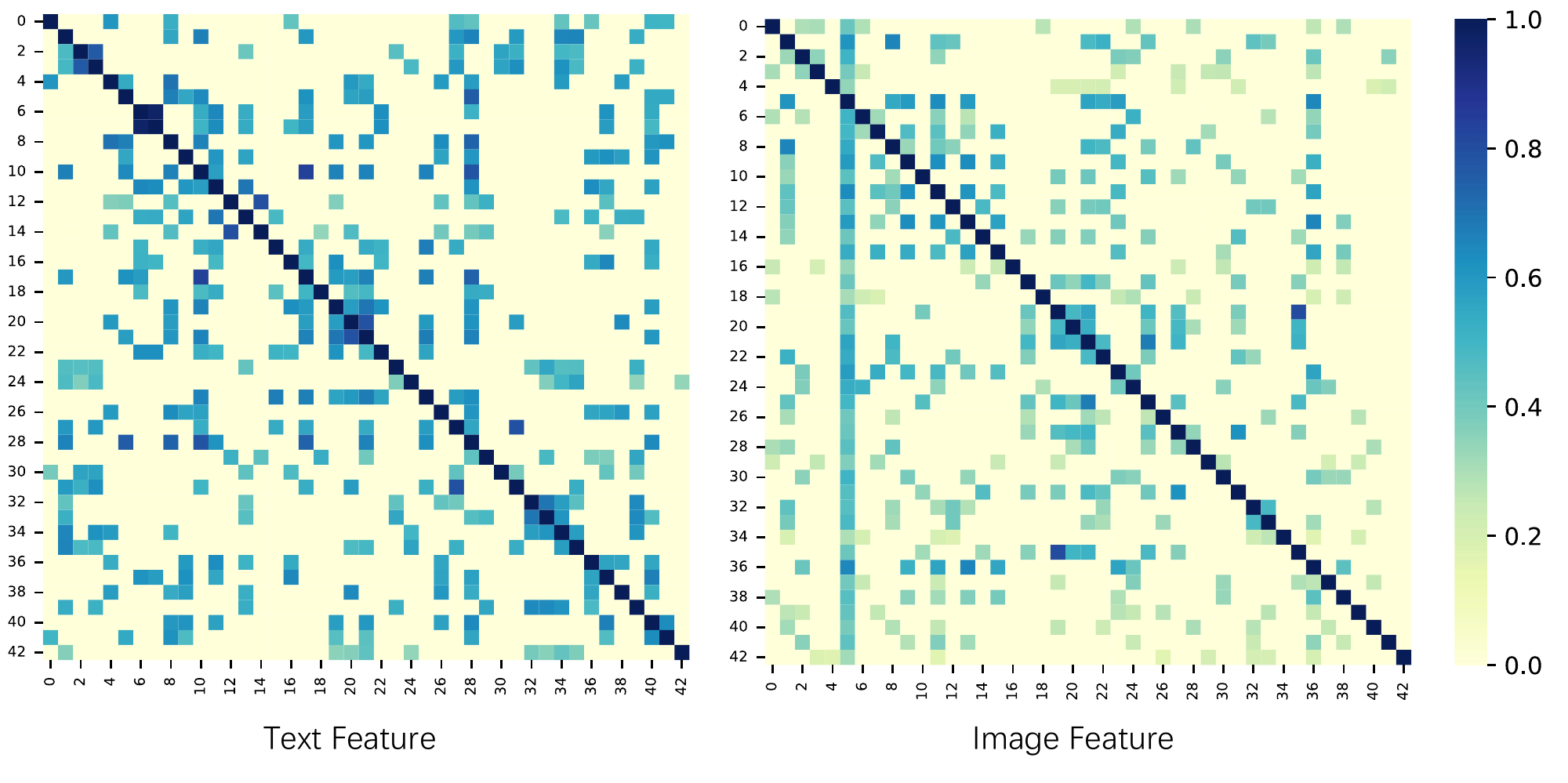}
    \vspace{-10pt}
    \caption{Semantic similarity of text and image features extracted by universal encoders. Each cell in the heat map represents the normalized similarity of two items, with horizontal and vertical coordinates denoting their mapped ID respectively. Darker hues indicate a higher similarity.}
    \label{fig:modal_heatmap}
    \vspace{-10pt}
\end{figure}

\begin{figure}
    \centering
    \includegraphics[scale=0.38]{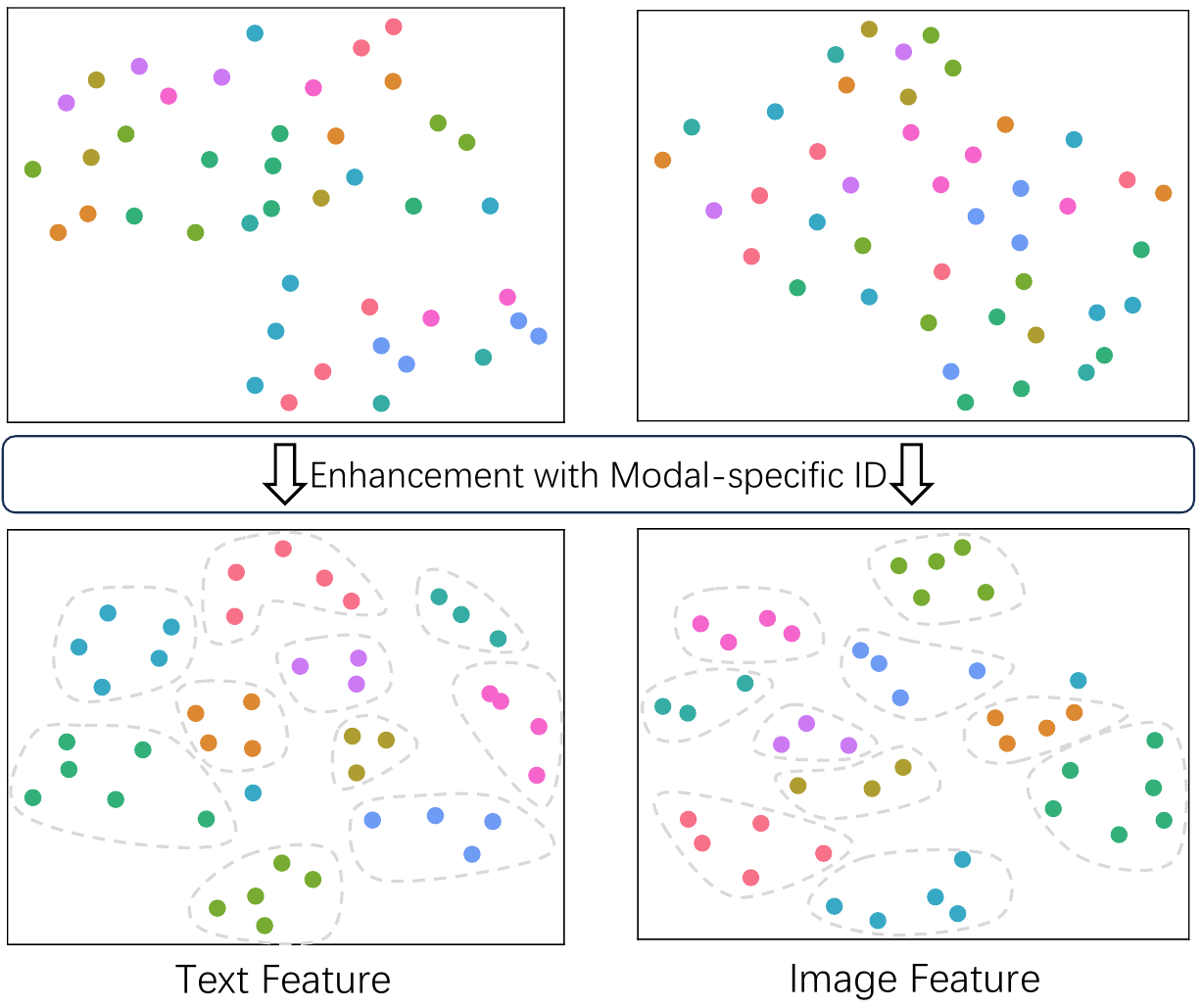}
    \vspace{-10pt}
    \caption{Distributions of image and text features before and after the enhancement. Items interacted with the same user are represented by the same color.}
    \label{fig:modal_scatter}
    \vspace{-10pt}
\end{figure}

Moreover, we visualize the item-item similarity matrix of pre-extracted multimodal features in the same way. We conduct analyses on both text and image features to ensure the generalizability of the results. As depicted in Figure~\ref{fig:modal_heatmap}, patterns of similarity distribution among different modalities are partial inconsistency, which indicates there are subtle differences in the semantics between different modalities. Therefore, it will lead to a performance drop by directly aligning or fusing different modalities when recommending. To mitigate this issue, we propose to enhance multimodal features with modal-specific ID embeddings (e.g. $tid$ and $vid$), thereby increasing their adaptability for recommendation. To confirm effectiveness, we plot the distributions of visual and textual features before and after enhancement (shown in Figure~\ref{fig:modal_scatter}), illustrating the effect of the enhancement on multimodal features in the context of recommendation systems. Our findings demonstrate that \emph{the inclusion of modal-specific ID embeddings effectively enhances these multi-modal features, leading to a significant improvement in the discriminability of their distributions}.

Consequently, to obtain optimal item representations, modal-specific ID embeddings are suggested to enhance multimodal features in terms of content and structure, respectively.

\section{Our IDSF Model}

Based on the findings in Section~\ref{sec:analysis}, we speculate that utilizing ID embeddings in a fine-grained manner can optimize item representations for better performance. In this section, we introduce our method that treats ID embeddings as subtle features of both content and structure, to enhance and fuse multimodal salient features. 

\subsection{Preliminary}
We represent the interaction data as a bipartite user-item graph $\mathcal{G}=\{(u,i)|u\in\mathcal{U},i\in\mathcal{I}\}$, where $\mathcal{U}$, and $\mathcal{I}$ denote the set of users and items, respectively. An edge $y_{ui}=1$ indicates a positive interaction between user $u$ and item $i$; otherwise $y_{ui}=0$. We denote the original features that have not been convoluted as $\textbf{e}_{i}^{m,(0)}$ and $\textbf{e}_{u}^{m,(0)}$ for a specific kind of information $m\in \mathcal{M}$ respectively, where $\mathcal{M}$ is the set of all kinds of information. Without sacrificing generality, we will refer to an item's textual $t$ and visual $v$ modalities as its salient feature and its associated item ID ($tid, vid$) as its subtle feature in this work. Future studies will also take into account the multimodal information provided by users. Specifically, we can obtain the definition as $\mathcal{M}=\{t, v, tid, vid\}$. In addition, the representations of higher-order items and users can be given as the $k$-layer graph convolution, denoted by $\textbf{e}_{i}^{m,(k)}$ and $\textbf{e}_{u}^{m,(k)}$, respectively. 
We divided the modal-specific bipartite graphs $\mathcal{G}_{t}$ and $\mathcal{G}_{v}$ from $\mathcal{G}$ in order to appropriately describe structural information on each modality individually.

\subsection{Model Overview}
\begin{figure*}[!htbp]
    \centering
    \includegraphics[width=6.5in]{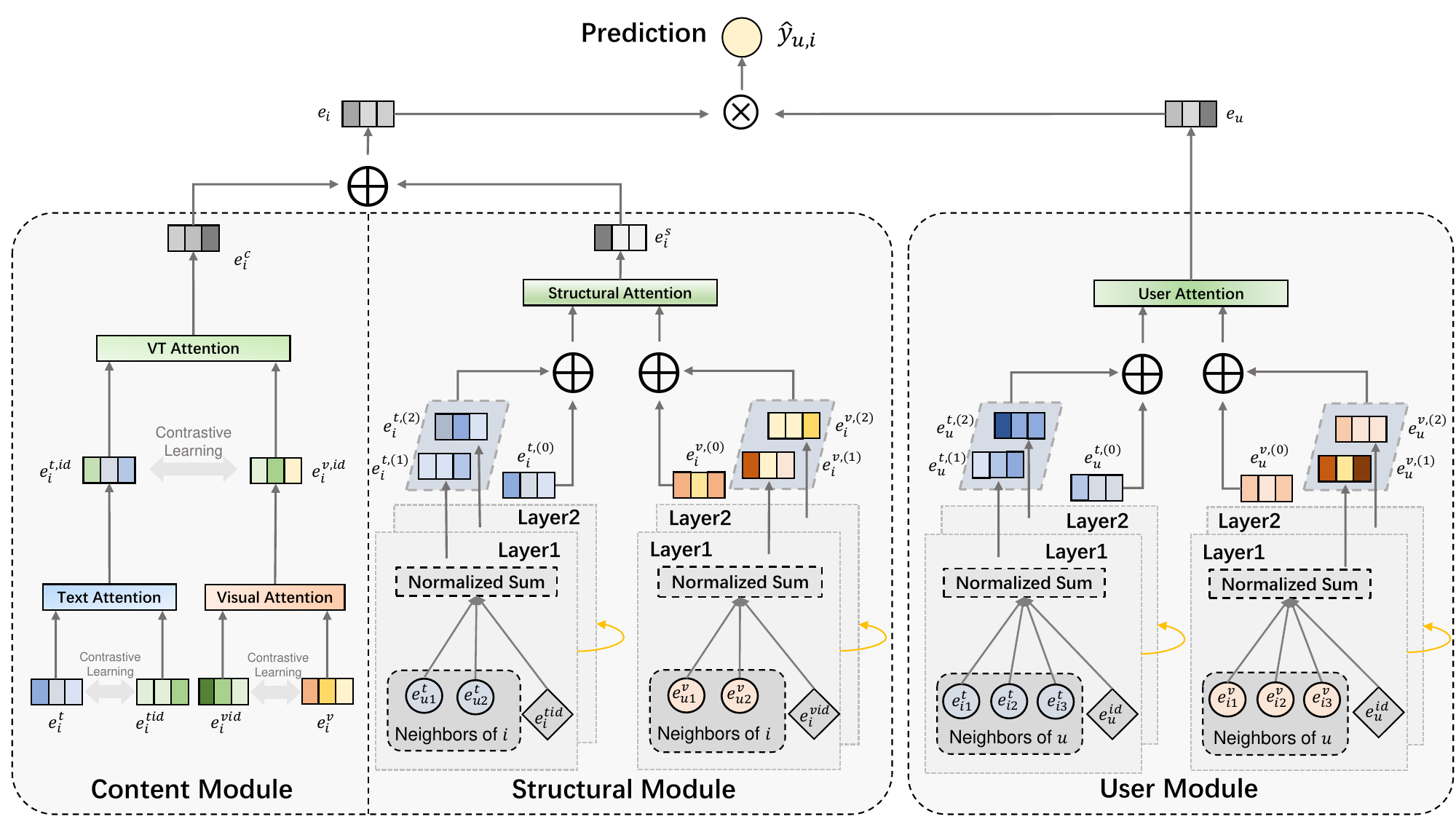}
    \vspace{-10pt}
    \caption{An illustration of our proposed IDSF framework. IDSF consists of three key modules: content representation module for items (left part), structural representation module for items (middle part), and users (right part). 
    Specifically, the item content module adopts a hierarchical attention mechanism to enhance multiple salient features (e.g., textual and visual in this figure) with subtle features in ID embeddings of items. The item- and user-structural modules design a lightweight modality-specific propagation mechanism to capture higher-order preference representations of items and users, respectively. The recommendation prediction generates a set of recommended items for a specific user.
    }
    \label{fig:model}
    \vspace{-10pt}
\end{figure*}

As illustrated in Figure~\ref{fig:model}, our IDSF model consists of three key components: \textbf{item content module}, \textbf{item structural module}, and \textbf{user structural module}.
Specifically, for the item content module, we aim to enhance the representation of each modality and their fusion by designing a hierarchical attention mechanism. That is, we improve the salient features (such as the textual and visual content in Figure~\ref{fig:model}) with the associated subtle features in item ID embeddings in each modality and fuse multiple modalities to generate the content representation $\textbf{e}_i^c$ of item $i$.
Then, in the item structural module, we adopt a similar idea to extract an item's high-order structural representation $\textbf{e}_i^s$ from various modalities of item $i$'s neighbors. We implement it by a multi-layer graph convolution in a lightweight way (textual modality on the left and visual modality on the right). The same process also holds for user structural module which obtains the user $u$'s structural representation $\textbf{e}_u^s$. 
Finally, we can obtain item $i$'s overall embedding by fusing the content representation $e_i^c$ with the structural representation $\textbf{e}_i^s$, and take user $u$'s structural representation $\textbf{e}_u^s$ as his/her final embedding. Prediction $\hat{y}_{ui}$ can be computed by the inner product of user $u$'s and item $i$'s embeddings. 

\subsection{Item Content Module}

Different from the existing works that mainly focus on salient features from multiple modalities, we further take into account subtle features from item ID embeddings. 
Specifically, we design a hierarchical attention mechanism in Section~\ref{subsubsec:fusion} to enhance salient content features (i.e., textual and visual) with their corresponding subtle features captured by modal-specific ID embeddings. Then, we introduce a contrastive loss to encourage the consistency among multi-modal information for better fusion in Section~\ref{subsubsec:cl}.

\subsubsection{Modality Enhancement and Fusion}
\label{subsubsec:fusion}

For each modality, we propose a hierarchical attention method to combine salient and subtle information. As shown in the left part of Figure~\ref{fig:model}, we first enhance the textual salient features $\textbf{e}^t_i$ with textual subtle features $\textbf{e}_i^{tid}$ in textual space through \emph{Text Attention} to get $\textbf{e}_i^{t^{\prime}}$. Similarly, we obtain visual features $\textbf{e}_i^{v^{\prime}}$ by \emph{Visual Attention}, which integrates the visual salient features $\textbf{e}^v_i$ with the visual subtle features $\textbf{e}_i^{vid}$ in the visual space. Then, we further fuse $\textbf{e}_i^{t^{\prime}}$ and $\textbf{e}_i^{v^{\prime}}$ through \emph{VT Attention} to get the comprehensive item content representation $\textbf{e}_i^{c}$.

All three attention mechanisms (Text Attention, Visual Attention, and VT Attention) follow the same procedure. 
We take the textual content and item ID in the textual space as an example. The modality fusion operation based on the attention mechanism is defined as follows:
\begin{equation}
    \textbf{e}^{t^{\prime}}_{i}=\sum_{m\in\{t,\; tid\}} \alpha^{m}_{i} \textbf{e}_{i}^{m},
    \label{eq:weight_sum}
\end{equation}
where $\textbf{e}_{i}^t$ represents the textual features extracted by sentence Bert~\cite{SentenceBERTSE}, $\textbf{e}_{i}^{tid}$ represents the learnable item ID embeddings in textual space, and $\alpha^{m}_{i}$ indicates the significance of each element $\textbf{e}_i^m$, calculated as follows:
\begin{equation}
    \alpha_{i}^m=\text{softmax}([\textbf{q}^{\top}\tanh(\mathbf{W}\textbf{e}^{m}_{i}+\textbf{b})], \; m \in \{t, tid\}),
    \label{eq:att}
\end{equation}
where $\textbf{q} \in \mathbb{R}^{d}$ denotes attention vector and $\textbf{W}\in\mathbb{R}^{d\times d}$, $b\in\mathbb{R}^{d}$ denote the trainable weight matrix and bias vector, respectively. We can compute $\textbf{e}_i^{v^{\prime}}$ and $\textbf{e}_i^{c}$ in the same way. In this way, we are able to enhance the salient (textual and visual) features with the subtle features of the item. 

\subsubsection{Contrastive Learning Constraints}
\label{subsubsec:cl}

In this paper, we construct a self-supervised contrastive learning task by maximizing the agreement between the item representations before and after fusion. 
Specifically, we separately maximize the degree of consistency between the representation of different information (e.g., $\textbf{e}_i^t$ and $\textbf{e}_i^{tid}$) and the fused representation (e.g., $\textbf{e}_i^{t^{\prime}}$) of the subtle and salient features. Similar comparisons are made between $\textbf{e}_i^v$, $\textbf{e}_i^{vid}$ and $\textbf{e}_i^{v^{\prime}}$, and also between $\textbf{e}_i^{t^{\prime}}$, $\textbf{e}_i^{v^{\prime}}$ and $\textbf{e}_i^{c}$. 

Here, we take the text content and item ID in the textual space as an example to formulate the contrastive learning loss as follows:
\begin{equation}
\resizebox{\linewidth}{!}{$
    \bar{\mathcal{L}}_{C}(\textbf{e}_{i}^t,\textbf{e}_{i}^{tid},\textbf{e}_{i}^{t^{\prime}} )\!=\!\!\!\\-\frac{1}{4|\mathcal{I}|}\sum\limits_{i\in\mathcal{I}}\sum\limits_{m\in\{t, \; tid\}}I(\textbf{e}_{i}^{m},\textbf{e}_{i}^{t^{\prime}})+ I(\textbf{e}_{i}^{t^{\prime}},\textbf{e}_{i}^{m}),$
    }
    \label{eq:contrastive_text}
\end{equation}
where $|\mathcal{I}|$ represents the number of items, and the mutual information $I(\cdot)$ between two representations can be mathematically expressed as:
\begin{equation}
\begin{split}
    I & (\textbf{e}_{i}^{m},\textbf{e}_{i}^{t^{\prime}}) =\log\frac{I_{ii}}{I_{ii}+I_{ij}}, \;\;\; m\in\{t, \; tid\}, \\
\end{split}
\label{eq:information1}
\end{equation}
where
\begin{equation}
\begin{split}
    & I_{ii} = \exp(f(\textbf{e}_{i}^{m},\textbf{e}_{i}^{t^{\prime}}))/\tau, \\
    & I_{ij} \!=\! \sum_{j\neq i}\left(\exp(f(\textbf{e}_{i}^{m},\textbf{e}_{j}^{t^{\prime}}))/\tau+\exp(f(\textbf{e}_{i}^{m},\textbf{e}_{j}^{m}))/\tau\right),
\end{split}
\label{eq:information2}
\end{equation}
where $i,j \in \mathcal{I}$, $\tau\in\mathbb{R}$ is a temperature hyper-parameter, $f(\cdot,\cdot)$ is the similarity function implemented by consine similarity.

The overall contrastive learning loss is given by:
\begin{equation}
\small
    \mathcal{L}_{C} = 
    \bar{\mathcal{L}}_{C}(\textbf{e}^t,\textbf{e}^{tid},\textbf{e}^{t^{\prime}}) + \bar{\mathcal{L}}_{C}(\textbf{e}^v,\textbf{e}^{vid},\textbf{e}^{v^{\prime}}) +
    \bar{\mathcal{L}}_{C}(\textbf{e}^{t^{\prime}},\textbf{e}^{v^{\prime}},\textbf{e}^{c})
    \label{eq:contrastive}
\end{equation}
In this way, we will get a more complete representation of a given item by aligning the representations of the various modalities with the contrastive learning loss.

\subsection{Item- and User- Structural Modules}
It's well known that the user-item interaction bipartite graph $\mathcal{G}$ created by recommender systems contains rich structural information. The interaction graph's higher-order connectedness depicts how preferences propagate across users and items. Graph neural network techniques have been widely proven to be effective in capturing higher-order graph structural information.
In this section, we will elaborate on a new method to aggregate the higher-order neighbor information by a lightweight graph convolution, and then we fuse the higher-order information of multiple modalities. 

Inspired by LightGCN~\cite{lightgcn} and MMGCN~\cite{mmgcn}, for a user (an item) node in a modality-specific interaction graph $\mathcal{G}_m$, we implement an aggregator via a weighted sum aggregator to gather information from its neighbors, and an enhancement through reserving the userID $\textbf{e}_u^{id}$ or modality-specific itemID $\textbf{e}_i^{id}$ embedding of the target node in a certain ratio $\gamma$. The lightweight message propagation mechanism of aggregated $k$-hop neighbors is expressed as follows:
\begin{equation}
    \textbf{e}_{i}^{m, (k)}=\sum_{u\in\mathcal{N}_i}\frac{1}{\sqrt{|\mathcal{N}_i|}\sqrt{|\mathcal{N}_u|}}\textbf{e}_{u}^{m,(k-1)} + \gamma\textbf{e}^{(m)id}_{i},
    \label{eq:eik}
\end{equation}
\begin{equation}
    \textbf{e}_{u}^{m,(k)}=\sum_{i\in\mathcal{N}_u}\frac{1}{\sqrt{|\mathcal{N}_u|}\sqrt{|\mathcal{N}_i|}}\textbf{e}_{i}^{m,(k-1)} + \gamma\textbf{e}^{id}_{u},
    \label{eq:euk}
\end{equation}
where $m \in \mathcal{M} = \{t, v\}$, $\textbf{e}_i^{m,(0)}=\textbf{e}_i^{m}$, $\textbf{e}_i^{(m)id}$ is the itemID that reuses the corresponding textual or visual space of the multimodal fusion module, i.e., $\textbf{e}_i^{(m)id}\in\{\textbf{e}_i^{vid},\textbf{e}_i^{tid}\}
$, 
$\textbf{e}_u^{id}$ represents the userID, which is a learnable variable initialized at random.
$\mathcal{N}_u$ and $\mathcal{N}_i$ denote the set of neighbors for user $u$ and item $i$ respectively, $\gamma$ is a hyper-parameter to control the amount of subtle information from the target node. The symmetric normalization term $\frac{1}{\sqrt{|\mathcal{N}_i|}\sqrt{|\mathcal{N}_u|}}$ follows the design of standard GCN to avoid the scale of embeddings increasing with graph convolution operations. 

After computing the high-order preference embedding at top-$K$ layers via Eq.~\ref{eq:eik} and Eq.~\ref{eq:euk}, we stack the preference embedding at each layer and take unweighted arithmetic mean to obtain the modal-specific representations of a user (an item). 
Finally, we derive the structural representations for user $u$ and item $i$ by an attentional combination of multimodal representations as Eq.~\ref{eq:eis} and Eq.~\ref{eq:eus} respectively,  defined by:
\begin{equation}
    \textbf{e}_{i}^{s}=\sum_{m \in \{t,v\}}\alpha_i^m \cdot (\frac{1}{K+1}\sum_{k=0}^{K}\textbf{e}_{i}^{m,(k)}),
    \label{eq:eis}
\end{equation}
\begin{equation}
    \textbf{e}_{u}^{s}=\sum_{m \in \{t,v\}}\alpha_i^m \cdot (\frac{1}{K+1}\sum_{k=0}^{K}\textbf{e}_{u}^{m,(k)}),
    \label{eq:eus}
\end{equation}
where $\alpha_i^m$ represents the importance of the modality $m$, and the calculation rules are shown in Eq.~\ref{eq:att}.

\subsection{Recommendation Prediction and Optimization}
The three above-mentioned key modules allow us to extract the content representation ($\textbf{e}_i^c$) for item $i$ as well as the high-order structure representation ($\textbf{e}_i^s$ and $\textbf{e}_u^s$) for item $i$ and user $u$, respectively. Therefore, we can predict user $u$'s preference for item $i$ as follows:
\begin{equation}
    \hat{y}_{ui}={\textbf{e}_{u}^s}^{\top} (\textbf{e}_{i}^{c}+\textbf{e}_{i}^{s}).
    \label{eq:y}
\end{equation}
where $\textbf{e}_i^c+\textbf{e}_i^s$ indicates the final representation of item $i$ by combining its content and structure\footnote{We do not introduce MLPs to map them since the structure and content features reside in the same latent space.}.

To train our model, we use Bayesian Personalized Ranking (BPR)~\cite{bpr} loss. BPR is a well-known pairwise loss that encourages the prediction of an observed user-item pair ($u,i$) to be higher than its unobserved counterparts ($u,j$):
\begin{equation}
    \mathcal{L}_{BPR}=-\sum_{u\in\mathcal{U}}\sum_{i\in\mathcal{N}_{u}}\sum_{j\notin\mathcal{N}_{u}}\ln\sigma(\hat{y}_{ui}-\hat{y}_{uj}),
    \label{eq:bpr}
\end{equation}
where $\sigma(\cdot)$ is the sigmoid function, $\mathcal{N}_{u}$ represents the set of all items that the user $u$ has interacted with.

We simultaneously optimize the multimodal contrastive loss and the BPR loss, that is, the overall objective function can be written as follows:
\begin{equation}
    \mathcal{L}=\mathcal{L}_{BPR}+\beta\mathcal{L}_{C}+\lambda||\Theta||_{2}^{2},
    \label{eq:loss}
\end{equation}
where $\beta$, $\lambda$ and $\Theta$ represent the strength of contrastive loss, the strengths of $L_{2}$ regularization, and the learnable parameters of the model, respectively.

Compared with other GCN-based multimodal recommendation algorithms, our method does not increase the time complexity of the algorithm but achieves better results which will be later verified by extensive experiments.

\section{Experiments and Analysis}
\label{sec:experiments}

\subsection{Experimental Settings}

\textbf{Datasets.} 
Three widely used Amazon\footnote{\footnotesize \url{http://jmcauley.ucsd.edu/data/amazon/links.html}} datasets are used in our experiments~\cite{Amazon}, including (a) \underline{Baby}, (b) \underline{Sports} and Outdoors, (c) \underline{Clothing}, Shoes and Jewelry. We are referred to as Baby, Sports, and Clothing in brief, respectively.  Table~\ref{tab:1} summarizes the statistics of three datasets. These datasets contain both visual and textual modalities of items other than user-item interactions. We extract visual features by CNN, and textual features by sentence-transformers~\cite{SentenceBERTSE} obtained by concatenating the title, description, category and brand of each item. All datasets are split into training, validation, and testing subsets with a ratio of 8:1:1. Three common metrics: $Precision@K$, $Recall@K$ and $NDCG@K$ are used for evaluation and $K$ is set to $10,20$ by default.

\begin{table}[t]
    \centering
    \caption{Statistics of the datasets.}
    \vspace{-10pt}
    \begin{tabular}{cccccc}
        \toprule 
        {Dataset} & {Modalities} & {\# Items} & {\# Users} & {\# Interactions} \\
        \hline 
        {Baby} & visual, textual & 7,050 & 19,445 & 139,110 \\
        {Sports} & visual, textual & 18,357 & 35,598 & 256,308 \\
        {Clothing} & visual, textual & 23,033 & 39,387 & 237,488 \\
        \bottomrule
    \end{tabular}
    \vspace{-10pt}
    \label{tab:1}
\end{table}

\begin{table*}[!htbp]
    \centering
    \caption{Performance of all comparison methods. Each row's second-best score is underlined and the top score is highlighted in bold. The final column indicates the percentage of performance improvement relative to the second best one. }
    \vspace{-10pt}
    \scalebox{1}{
    \begin{tabular}{c|l|ccc|cccccc|c|c}
    \hline
        \multirow{2}{*}{Dataset} & \multirow{2}{*}{Metric($\times$100$\%$)} & \multicolumn{3}{|c|}{\textbf{CF Methods}} & \multicolumn{7}{|c|}{\textbf{Multi-modal Methods}} & \multirow{2}{*}{Improve} \\ \cline{3-12}
        \multicolumn{1}{c|}{} & \multicolumn{1}{|c|}{} & \multicolumn{1}{|c}{BPR} & NGCF & LightGCN & VBPR & MMGCN & GRCN & SLMREC & MICRO & BM3 & \textbf{IDSF} & ~ \\ 
        \hline
        \multicolumn{1}{c|}{\multirow{6}[1]{*}{Baby}} & Recall@10 & 2.935 & 3.567 & 4.509 & 3.103 & 3.861 & 4.443 & 5.233 & \underline{5.807} & 4.481 & \textbf{6.143} & 5.8\% \\ 
        ~ & Precision@10 & 0.308 & 0.378 & 0.478 & 0.332 & 0.408 & 0.473 & 0.585 & \underline{0.608} & 0.470 & \textbf{0.645} & 6.0\% \\ 
        ~ & NDCG@10 & 1.588 & 1.964 & 2.498 & 1.749 & 1.996 & 2.392 & 2.885 & \underline{3.218} & 2.343 & \textbf{3.431} & 6.6\% \\ 
        \cline{2-12}
        ~ & Recall@20 & 4.566 & 5.751 & 7.248 & 4.863 & 6.240 & 7.895 & 7.775 & \underline{8.884} & 6.688 & \textbf{9.505} & 6.9\% \\ 
        ~ & Precision@20 & 0.243 & 0.307 & 0.383 & 0.262 & 0.332 & 0.422 & 0.445 & \underline{0.466} & 0.354 & \textbf{0.499} & 7.1\% \\ 
        ~ & NDCG@20 & 2.033 & 2.552 & 3.212 & 2.330 & 2.605 & 3.552 & 3.541 & \underline{4.024} & 2.931 & \textbf{4.311} & 7.1\% \\ 
        \hline
        \multicolumn{1}{c|}{\multirow{6}[1]{*}{Sports}} & Recall@10 & 2.977 & 4.925 & 5.786 & 3.560 & 3.352 & 5.879 & 6.559 & \underline{6.671} & 5.947 & \textbf{7.001} & 5.0\% \\ 
        ~ & Precision@10 & 0.318 & 0.521 & 0.611 & 0.382 & 0.355 & 0.627 & 0.693 & \underline{0.701} & 0.659 & \textbf{0.734} & 4.7\% \\ 
        ~ & NDCG@10 & 1.712 & 2.693 & 3.376 & 2.086 & 1.763 & 3.310 & 3.726 & \underline{3.728} & 3.158 & \textbf{4.006} & 7.5\% \\ 
        \cline{2-12}
        ~ & Recall@20 & 4.386 & 7.471 & 8.502 & 5.211 & 5.466 & 8.752 & 9.834 & \underline{9.951} & 9.153 & \textbf{10.41} & 4.6\% \\ 
        ~ & Precision@20 & 0.236 & 0.397 & 0.451 & 0.281 & 0.291 & 0.467 & 0.518 & \underline{0.525} & 0.517 & \textbf{0.548} & 4.4\% \\ 
        ~ & NDCG@20 & 2.095 & 3.371 & 4.102 & 2.533 & 2.284 & 4.078 & 4.588 & \underline{4.599} & 3.978 & \textbf{4.914} & 6.8\% \\ 
        \hline
        \multicolumn{1}{c|}{\multirow{6}[1]{*}{Clothing}} & Recall@10 & 1.322 & 2.693 & 3.684 & 2.199 & 1.996 & 4.006 & 4.964 & \underline{5.047} & 4.144 & \textbf{5.381} & 6.6\% \\ 
        ~ & Precision@10 & 0.135 & 0.273 & 0.373 & 0.223 & 0.202 & 0.407 & 0.502 & \underline{0.436} & 0.656 & \textbf{0.545} & 6.5\% \\ 
        ~ & NDCG@10 & 0.715 & 1.441 & 2.032 & 1.236 & 1.014 & 2.164 & 2.711 & \underline{2.226} & 3.152 & \textbf{2.966} & 6.5\% \\ 
        \cline{2-12}
        ~ & Recall@20 & 1.938 & 4.095 & 5.407 & 3.175 & 3.231 & 6.193 & 7.471 & \underline{7.674} & 6.388 & \textbf{7.915} & 3.1\% \\ 
        ~ & Precision@20 & 0.099 & 0.208 & 0.274 & 0.161 & 0.164 & 0.315 & 0.378 & \underline{0.388} & 0.333 & \textbf{0.401} & 3.2\% \\ 
        ~ & NDCG@20 & 0.875 & 1.801 & 2.471 & 1.456 & 1.325 & 2.721 & 3.347 & \underline{3.451} & 2.834 & \textbf{3.611} & 4.6\% \\ 
        \hline
    \end{tabular}
    }
    \label{tab:2}
\end{table*}

\textbf{Baselines.}
We compare IDSF with nine competing methods, which can be divided into two categories: \emph{CF methods} and \emph{multimodal methods}. The first category contains three classic recommendation methods that merely generate recommendations based on user-item interactions with no consideration of modality information, including {BPR}~\cite{bpr}, {NGCF}~\cite{ngcf}, and {LightGCN}~\cite{lightgcn}.
The second category of our comparison methods that take into account additional modality information for item representations before recommendation, including {VBPR}~\cite{vbpr}, {MMGCN}~\cite{mmgcn}, {GRCN}~\cite{grcn}, {SLMREC}~\cite{slmrec}, {MICRO}\footnote{MICRO is an extender version of LATTICE\cite{lattice}, so we omitted the performance of LATTICE.}~\cite{micro} and {BM3}~\cite{bm3}. Details about baselines are provided in Appendix \ref{app:baselines}.

\textbf{Hyperparameters.}
For a fair comparison, we set the embedding dimension to 128, batch size to 1024, and initialized all model parameters with Xavier initializer~\cite{Xavier} which is optimized by Adam optimizer~\cite{AdamAM}. Other hyper-parameters are determined via grid search on the validation set. Finally, the learning rate is set to $0.0005$, the coefficient of $\ell_{2}$ is set to $10^{-4}$, the temperature parameter $\tau$ is set to 0.5, the coefficient $\beta$ used to control the effect of contrastive multimodal fusion task is set to 0.3 for dataset {Baby}, 1.0 for {Sports} and {Clothing}, the coefficient $\gamma$ used to control the amount of information from cross-modal representations of the target node is set to 0.3 for dataset {Baby}, 0.3 for {Sports} and 1.0 for {Clothing}. We set the number of GCN layers to $K=2$ the same as LightGCN~\cite{lightgcn}. Besides, we adopt an early-stop strategy if Recall@20 on the validation set no longer increases after $5$ epochs to avoid overfitting and follow the original settings of comparison methods to achieve the best performance. Due to space limitations, we include the analysis of hyperparameters in the Appendix~\ref{app:hyper}.

\begin{table*}[!htbp]
    \centering
    \caption{Performance of different variants of our IDSF.}
    \vspace{-10pt}
    \scalebox{.76}{
    \begin{tabular}{c|cccccc|cccccc|cccccc}
    \toprule
        Dataset & \multicolumn{6}{|c|}{Baby} & \multicolumn{6}{|c|}{Sports} & \multicolumn{6}{|c}{Clothing} \\ 
        \midrule
        Metric($\times$100$\%$) & R@10 & P@10 & N@10 & R@20 & P@20 & N@20 & R@10 & P@10 & N@10 & R@20 & P@20 & N@20 & R@10 & P@10 & N@10 & R@20 & P@20 & N@20 \\
        \midrule
        w/o content & 5.211 & 0.551 & 2.865 & 8.258 & 0.437 & 3.671 & 5.539 & 0.584 & 3.071 & 8.383 & 0.444 & 3.831 & 3.963 & 0.401 & 2.181 & 5.899 & 0.298 & 2.673 \\ 
        content w/o contrast & 5.551 & 0.582 & 3.089 & 8.654 & 0.455 & 3.886 & 5.754 & 0.608 & 3.181 & 8.763 & 0.464 & 3.981 & 4.175 & 0.422 & 2.285 & 6.229 & 0.315 & 2.811 \\ 
        content w/o ID & 5.923 & 0.625 & 3.236 & 9.276 & 0.488 & 4.114 & 6.857 & 0.721 & 3.897 & 10.21 & 0.538 & 4.783 & 4.927 & 0.499 & 2.668 & 7.298 & 0.369 & 3.271 \\ 
        structure w/o ID & 5.629 & 0.591 & 3.152 & 8.611 & 0.452 & 3.931 & 6.498 & 0.682 & 3.573 & 9.653 & 0.507 & 4.403 & 5.151 & 0.521 & 2.781 & 7.698 & 0.391 & 3.433 \\ 
        \midrule
        IDSF & 6.143 & 0.645 & 3.431 & 9.505 & 0.499 & 4.311 & 7.001 & 0.734 & 4.006 & 10.41 & 0.548 & 4.914 & 5.381 & 0.545 & 2.966 & 7.915 & 0.401 & 3.611 \\
        \bottomrule
    \end{tabular}
    }
    \label{tab:3}
\end{table*}

\begin{table*}[!htbp]
    \centering
    \caption{Performance comparison over enhanced modalities and original modalities. Where, \textbf{original visual}, \textbf{original textual} and \textbf{original fused} modalities represent that the content and structure features are obtained by using the original modalities without the subtle features, while \textbf{enhanced visual}, \textbf{enhanced textual} and \textbf{enhanced fused} modalities mean that the original features are enhanced by the subtle features, respectively.}
    \vspace{-10pt}
    \scalebox{.78}{
    \begin{tabular}{c|cccccc|cccccc|cccccc}
    \toprule
        Dataset & \multicolumn{6}{|c|}{Baby} & \multicolumn{6}{|c|}{Sports} & \multicolumn{6}{|c}{Clothing} \\ 
        \midrule
        Metric($\times$100$\%$) & R@10 & P@10 & N@10 & R@20 & P@20 & N@20 & R@10 & P@10 & N@10 & R@20 & P@20 & N@20 & R@10 & P@10 & N@10 & R@20 & P@20 & N@20 \\
        \midrule
        original visual & 3.068 & 0.327 & 1.607 & 5.356 & 0.285 & 2.217 & 2.046 & 0.214 & 1.012 & 3.523 & 0.186 & 1.406 & 1.284 & 0.131 & 0.611 & 2.361 & 0.119 & 0.882 \\ 
        original textual & 3.164 & 0.333 & 1.735 & 5.628 & 0.292 & 2.392 & 2.075 & 0.218 & 1.046 & 3.724 & 0.197 & 1.482 & 1.378 & 0.139 & 0.942 & 2.493 & 0.127 & 0.942 \\ 
        original fused & 5.048 & 0.529 & 2.619 & 8.221 & 0.431 & 3.451 & 5.983 & 0.628 & 3.287 & 9.113 & 0.479 & 4.111 & 4.614 & 0.466 & 2.437 & 7.203 & 0.365 & 3.097 \\
        \hline
        enhanced visual & 5.639 & 0.593 & 3.112 & 8.866 & 0.466 & 3.958 & 6.831 & 0.719 & 3.861 & 10.11 & 0.535 & 4.738 & 4.798 & 0.485 & 2.627 & 7.041 & 0.357 & 3.199 \\
        enhanced textual & 5.956 & 0.625 & 3.224 & 9.251 & 0.481 & 4.085 & 6.925 & 0.729 & 3.911 & 10.32 & 0.539 & 4.806 & 5.224 & 0.527 & 2.826 & 7.799 & 0.394 & 3.508 \\
        enhanced fused & 6.143 & 0.645 & 3.431 & 9.505 & 0.499 & 4.311 & 7.001 & 0.734 & 4.006 & 10.41 & 0.548 & 4.914 & 5.381 & 0.545 & 2.966 & 7.915 & 0.401 & 3.611 \\
        \bottomrule
    \end{tabular}
    }
    \label{tab:enhance}
\end{table*}

\vspace{-5pt}
\subsection{Performance Comparison}
The comparative results are summarized in Table \ref{tab:2}, from which we can find that our proposed IDSF method outperforms both CF methods and multimodal methods. Generally, multimodal methods perform better than CF methods, implying the value of extracting salient features from multimodal information for item representation. 

For the CF methods without considering multimodal information, NGCF performs better than BPR since the former method captures high-order structural features from the interactions between users and items, implying that it is significant to model structural features through graphs. In comparison, LightGCN performs the best in CF methods thanks to its adoption of an optimized lightweight aggregation method (rather than the aggregation in NGCF). This aggregation method obtains more precise structural features, demonstrating that optimized structural features can help improve recommendation performance.

Among the multimodal methods, VBPR obtains the worst performance due to the mere consideration of visual content features without structural features. Nevertheless, the fact that VBPR beats BPR to a certain extent implies the effectiveness of content features. 
MMGCN performs better by extracting additional structural features of textual and visual modalities. Meanwhile, we find that MMGCN outperforms NGCF under the same convolution method, leading to the conclusion that it is important to improve the structural features with multimodal information. 
Furthermore, GRCN refines the structural features by identifying the false-positive feedback and fixing it, whereby better performance is obtained. It is suggested to strengthen the structural features in multimodal recommendation. 
However, LightGCN achieves the same performance as GRCN only by optimizing the aggregation method, emphasizing the effect of the optimized aggregation in improving the structural features. 
MICRO gain comparable sub-optimal performance, owing to their fine-grained multimodal fusion with contrastive learning. The significance of contrastive learning in multimodal fusion is self-evident.

Finally, our proposed IDSF method beats the second-best comparison method in terms of three metrics by around 5.8-7.1\% on dataset Baby, 4.4-7.5\% on Sports, and 3.1-6.6\% on Clothing, respectively. We attribute these important improvements mainly to the fine-grained employment of ID embeddings for multimodal recommendation, which can provide valuable subtle features (extracted from ID embeddings) to complement with salient features (from multi-modal information). The complementation can help better learn the representations of items from the perspectives of both content and structure.

\subsection{Ablation Studies}
To explore the effects of content features, contrastive learning and ID embeddings, we compare the results on four variants: \textit{w/o content}, which discards item content features, \textit{content w/o contrast} which omits contrastive loss, \textit{content w/o ID} which skips fusing subtle features in content features and \textit{structure w/o ID} which leaves out ID embeddings when obtaining structural features. Table~\ref{tab:3} summarizes the performance of different variants of IDSF, from which we have the following observations.

Without the support of content features, the performance of \textit{IDSF w/o content} significantly decreases, indicating that the content features are essential, which can most explicitly reveal multimodal information of items.
Comparing \textit{content w/o contrast} and \textit{content w/o ID}, we can observe that it is necessary to draw contrastive learning into the content enhancement with subtle features. If there is no contrastive learning, ID embeddings tend to generate more noisy subtle features which cause performance degradation.
\textit{Content w/o ID} surpasses \textit{structure w/o ID} since structural features are more sensitive to ID embeddings. Therefore, for structure features, we control the amount of ID information in the enhancement contained in convolution operation with hyperparameters.

Finally, IDSF outperforms all variants on three datasets across these three evaluation metrics, further validating the significance of the enhancement with subtle features. It shows that subtle features can complement salient features and increase the appropriateness of content and structural representation to produce more accurate recommendations.

\vspace{-10pt}
\subsection{The Contribution of Enhancement in Modality Missing}
The modality missing problem remains a prominent challenge in multimodal recommendation. ID embeddings provide rich subtle information of content and structure, which can be exploited to enhance the expressiveness of salient features. In this subsection, we conduct incomplete modality experiments, aiming to explore the contribution of the proposed enhancement in the case of missing modalities. Tabel~\ref{tab:enhance} reports the performance comparison over enhancing modalities with subtle features and original modalities.

According to the performance of the original and enhanced single-modality, it is observed that the proposed enhancement is quite effective in making important improvements. Meanwhile, the fusing scenario of enhanced textual and visual features outperforms that of original features, demonstrating the value of subtle features for better representation of items. In other words, our proposed enhancement can alleviate the degradation of recommendation performance when certain modalities are missing.

Moreover, textual modality is more useful than visual modality in general to model items. It can be explained by the fact that textual modality is able to provide finer-grained attributes such as categories and descriptions of items.
The performances of utilizing multiple enhanced features are significantly better than those of ones with a single modality. The multimodal combination makes item representations more comprehensive and reaches better performance. The enhanced modalities follow the same trend, which implies that our proposed enhancement by subtle features extracted from ID embeddings does not interfere with the advantages of multiple modalities over a single modality.

\vspace{-5pt}
\section{Related Work}

\subsection{ID-based Recommendation}

ID embeddings contain abundant and important internal characteristics and are essential in the existing recommendation literature\cite{IDvsModal}. For non-sequential tasks, recommendation models evolve from the early item-item collaborative filtering~\cite{CF}, various matrix factorization-based approaches~\cite{mf,dtmf,TrustSVDCF,ncf, DeepFM}, to graph-based methods~\cite{grcn,lightgcn,ultragcn,Wu2021SelfsupervisedGL, HypergraphContrastiveCF}, taking user-item pairs as input to predict matching scores between users and items. For sequential tasks, recommendation models with a sequential prediction model, such as RNN, LSTM, GRU and Transformer as the backbone, taking a user's historical sequence of items as input to generate the next interactions~\cite{SASRec,Cui2020MVRNNAM,session_based_GNN,BERT4Rec,UniformBetter}. These recommendation methods treat user/item ID as a complete item representation to conduct predictions.

\subsection{ID in Multimodal Recommendation}

Multimodal recommendations leverage ID embedding as a component, yet they do not assign significant importance to this particular technique. Matrix factorization-based approaches incorporate ID embeddings derived from historical interactions as content features for modeling user preferences. In this context, multimodal features are concatenated with ID embeddings as side information to enrich item and user representations. For instance, VBPR~\cite{vbpr} adds the score obtained by the inner product of item image features and user visual preference in prediction calculation as an extension of ID embeddings. DeepStyle~\cite{deepstyle} employs a shared user latent factor to interact with both image features and item ID embeddings. ACF~\cite{acf} utilizes attention mechanisms to encode user preferences while incorporating user and item IDs as latent factors, which are subsequently aggregated.

In some GCN-based methods~\cite{mmgcn,grcn}, ID embeddings are integrated into modality-specific structural features as a unique cross-modal global feature. These methods create modality-aware interaction graphs and perform information propagation and fusion to aggregate neighbors' information for target nodes, and ID embeddings are considered as modality connections. SLMREC~\cite{slmrec} realizes the underlying semantic role of ID embeddings in recommendation, and treat ID embeddings as a general modality, performing graph convolution on "ID-modality".

Moreover, a few works use ID embeddings similarly just as in the conventional recommendation. For example, MICRO~\cite{micro}, LATTICE~\cite{lattice}, BM3\cite{bm3} and FREEDOM\cite{freedom} use ID embeddings only for the needs of collaborative filtering tasks, and take multimodal module as an auxiliary task without considering the connections between ID embeddings and multi-modalities. CMFB~\cite{cmfb}, DualGNN~\cite{DualGNN}, and HUIGN~\cite{HUIGN} abandon ID and simply concatenate multimodal content to replace item IDs for matrix factorization or graph convolution.

Recently, Xiao et al.~\cite{abstracttodetails} propose a generative multimodal fusion framework (GMMF) for the CTR prediction task, which improves multimodal features by generating new visual and text representations by a Difference Set network (DSN). It maps item ID into modal space and treats it as a "special modality" to model the difference and connections between modalities as PAMD~\cite{PAMD}.

Although the usage of ID varies, they all regard ID as a whole without explanation. Contrarily, we contend that a more nuanced exploration of ID utilization is warranted. In our method, we step further by discovering underlying information and fine-grained utilization of ID to model item representations more comprehensively.

\vspace{-5pt}
\section{Conclusion and Future Work}

In this paper, we revisit the significance of ID embeddings and conduct analyses for ID embeddings and multimodal features in the context of multimodal recommendation. Leveraging the subtle features within ID embeddings and recognizing the distinctions between modalities, we propose a novel multimodal recommendation method called IDSF to use subtle content and structural features in ID embeddings effectively, reaching desirable performance.

For future work, we intend to further consider the subtle features in sequential recommendation and find appropriate methods to integrate them. For users, exposed content salient features are unavailable in some datasets so we are looking for this information from users' interactive behaviors like historical reviews, etc. Moreover, in an era dominated by large-scale models, the exploration of ID embeddings in pre-training and fine-tuning is worth further investigation.



\bibliographystyle{ACM-Reference-Format}
\bibliography{reference}



\clearpage

\appendix

\section{Experimental Details}
\label{app:detail}

In Section \ref{sec:analysis}, we conducted a visualization analysis of the pre-trained ID embeddings (referred to as IDs) and multimodal features. To ensure the generalizability of our experiments, we randomly selected 10 users from the training set and created a sample set that included all the items they interacted with for analysis. After sampling, we checked for any overlap in the interaction items between users to avoid excessive complexity in the experiment.

For the scatter plots (Figure \ref{fig:id_scatter} and Figure \ref{fig:modal_scatter}), we used UMAP~\cite{UMAP} to reduce the dimensions of the IDs to 2-D and plotted each ID using the \textit{matplotlib} package. Points that are closer in the plot indicate that they are also closer in the latent space, revealing structural similarity. Points of the same color (i.e., item IDs interacted with by the same user) that are closer together indicate that the IDs can reflect structural similarity.

For the heatmaps (Figure \ref{fig:id_heatmap} and Figure \ref{fig:modal_heatmap}), we directly computed the cosine similarity between each pair of item IDs, resulting in a similarity matrix of size $43\times43$ (where 43 is the number of items), which was visualized as a heatmap. To enhance clarity, we only retained the top 10 values of each row in the similarity matrix, setting the rest to 0. The heatmaps directly display the semantic similarity between IDs, and it is evident that IDs of items that interacted with the same user exhibit higher semantic similarity, indicating that IDs can reflect content similarity.

Based on these findings, we believe that the IDs can be categorized as \emph{content features} and \emph{structure features}.

\section{Baselines}
\label{app:baselines}

We compare IDSF with nine competing methods, which can be divided into two categories: \textbf{CF methods} and \textbf{multimodal methods}. The first category contains three classic recommendation methods that merely generate recommenation based on user-item interactions with no consideration of modality information. 
\begin{itemize}
    \item \textbf{BPR} \cite{bpr} is a classic item ranking method built upon the assumption that a user prefers an interacted item to an unknown one. 
    \item \textbf{NGCF} \cite{ngcf} explicitly models the user-item interactions by a bipartite graph, uses graph convolution operation to learn the topology, and effectively harvests the high-order connectivity and collaborative signals for item recommendation.
    \item \textbf{LightGCN} \cite{lightgcn} abandons the use of feature transformation and nonlinear activation, and only retains the most important neighbor aggregation module in GCNs for collaborative filtering.
\end{itemize}
\textbf{Multimodal models} are the second category of our comparison methods that take into account additional modality information for item recommendation. 
\begin{itemize}
    \item \textbf{VBPR} \cite{vbpr} integrates the visual features and ID embedding of each item as its representations for the matrix factorization.
    \item \textbf{MMGCN} \cite{mmgcn} is one of the state-of-the-art multimodal recommendation models, which constructs a graph convolution network from multiple modalities to improve the structural representation.
    \item \textbf{GRCN} \cite{grcn} is also one of the state-of-the-art multimodal recommendation methods. It refines the user-item interaction graph by identifying the false-positive feedback and fixing it.
    \item \textbf{SLMREC} \cite{slmrec} incorporates contrastive learning into modal-specific graph neural network to promote multimodal fusion.
    \item \textbf{MICRO} \cite{micro} mines latent item relations and conducts fine-grained multimodal fusion with contrastive learning before collaborative filtering as a auxiliary task. 
    \item \textbf{BM3} \cite{bm3} is a self-supervised multi-modal recommendation model, which requires neither augmentations from auxiliary graphs nor negative samples.
\end{itemize}

\section{Hyperparameters Analysis}
\label{app:hyper}

We do sensitivity analysis with various hyper-parameters on the graph convolution and the contrastive task in this paragraph because they are crucial components of our strategy. We examine IDSF performance in relation to various $\gamma$ and $\beta$ values. The $\gamma$ dictates how much multimodal subtle information from the target node should be retained. The performance impact of the contrastive task magnitude $\beta$ is then covered.
\begin{figure*}[!hb]
   \vspace{-10pt}
	\centering
        \includegraphics[scale=0.25]{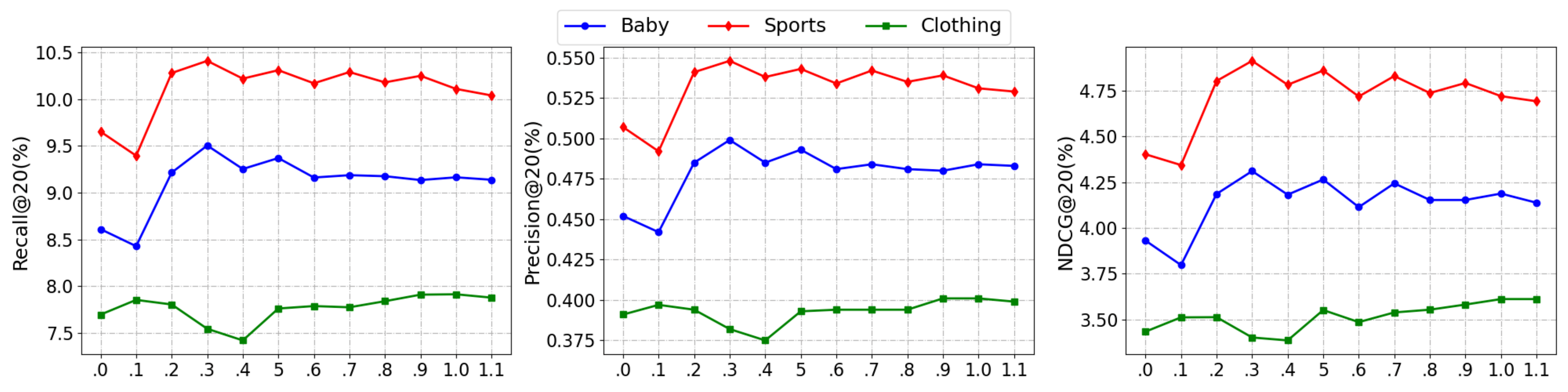}
        \vspace{-10pt}
	\caption{Performance evaluation across various gamma($\gamma$) values.}
	\label{fig:gamma}
 \vspace{-10pt}
\end{figure*}
\begin{figure*}[b]
	\centering
        \includegraphics[scale=0.25]{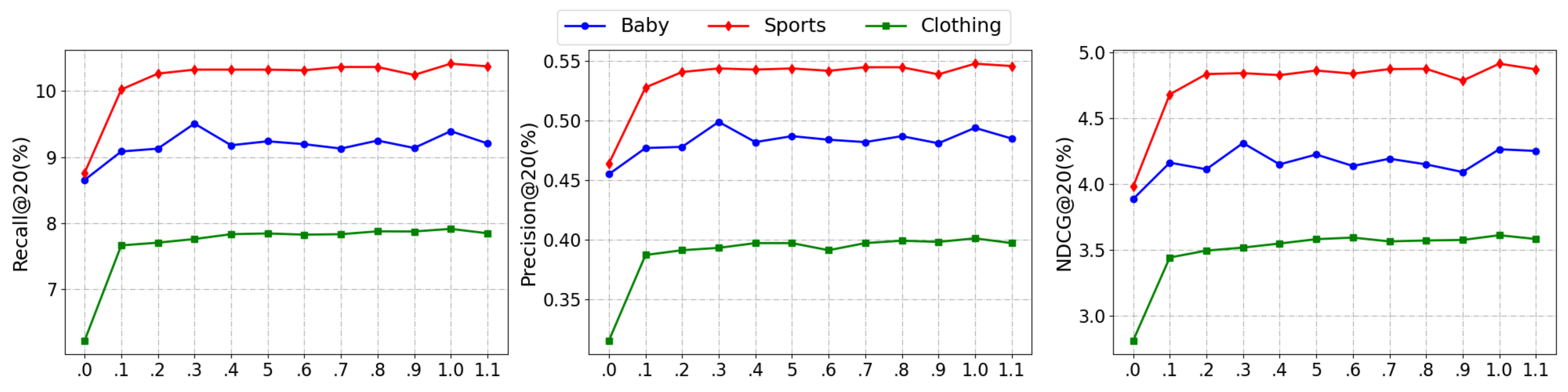}
        \vspace{-10pt}
	\caption{Performance evaluation across various beta($\beta$) values.}
	\label{fig:beta}
\end{figure*}

Figure~\ref{fig:gamma} reports the results of performance comparison. $\gamma=0$ means no subtle features reserved while doing graph convolution since it collects just the linked neighbors and ignores the self-connections (i.e., structure w/o ID). Moreover, we can observe the following:
\begin{itemize}
    \item When $\gamma$ is adjusted to 0.3, 0.3, and 1.0, respectively, performance on Baby, Sport, and Clothing yields the best results, which validates the significance of modal-specific subtle features in structural representations. Appropriate $\gamma$ can obtain better item representations by aggregating subtle and salient structural features which boost the recommendation performance.
    \item When the performance on three datasets reaches the peak, the corresponding $\gamma$ are not the same because the importance of salient features on revealing items attributes varies for different types of items. The improving trend declines when $\gamma$ exceeds the best value since the excessive proportion of subtle features will cover the effect of the salient features.
    \item Furthermore, given that it is sparse than other datasets that require more multimodal items to provide better recommendations, performance on \textit{Clothing} improves as $\gamma$ increases. 
\end{itemize}

The effect of various coefficients $\beta$ on performance is seen in Figure~\ref{fig:beta}. $\beta=0$ denotes IDSF w/o contrast, which discards the contrastive learning task. We can observe that
\begin{itemize}
    \item The performances on all datasets first improve as beta increases and are always better than $\beta=0$. The primary task achieves improvements when jointly optimized with the contrastive self-supervised auxiliary task when with a small $\beta$. This implies that it is important to take multimodal fusion's semantic consistency into account.
    \item When $\beta$ continues to increase, it starts to drop, indicating that $\beta$ interferes with the main work of making recommendations and the gain brought by the self-supervised task could be counteracted when $\beta$ is more than the weight of the BPR task.
    \item Overall, there are no apparent sharp rise and falls when $\beta\neq0$, which indicates that our methods is not that sensitive to the selection of the ratio of auxiliary task.
\end{itemize}

\end{document}